\documentclass[proceedings]{stacs}
\stacsheading{2010}{71-82}{Nancy, France}
 \firstpageno{71}

\theoremstyle{plain}
\theoremstyle{definition}

\usepackage{amsmath, amsthm, amssymb}
\numberwithin{equation}{section}
\DeclareMathOperator{\li}{li}
\DeclareMathOperator{\sym}{Sym}
\DeclareMathOperator{\aut}{Aut}
\DeclareMathOperator{\agl}{AGL}
\begin{document}

\newcommand{\ie}{i.\,e.}
\newcommand{\F}{\mathbb{F}}
\newcommand{\Z}{\mathbb{Z}}
\newcommand{\vf}{\varphi}
\newcommand{\calK}{{\mathcal K}}

\title[Evasiveness and the Distribution of Prime Numbers]{Evasiveness 
and the Distribution of Prime Numbers}

\author[lab1, lab3]{L. Babai}{L\'aszl\'o Babai}
\address[lab1]{University of Chicago, Chicago, IL, USA.
  }  


\author[lab2]{A. Banerjee}{Anandam Banerjee}
\address[lab2]{Northeastern University, Boston, MA, USA.}  

\author[lab1]{R. Kulkarni}{Raghav Kulkarni}

\author[lab1]{V. Naik}{Vipul Naik}






\begin{abstract}
A Boolean function on $N$ variables is called \emph{evasive}
if its decision-tree complexity is $N$.   A sequence $B_n$ of
Boolean functions is \emph{eventually evasive} if $B_n$ is
evasive for all sufficiently large $n$.

We confirm the eventual
evasiveness of several classes of monotone graph properties
under widely accepted number theoretic hypotheses.  In particular we show
that Chowla's conjecture on Dirichlet primes implies
that (a) for any graph $H$, ``forbidden subgraph $H$'' is eventually evasive
and (b) all nontrivial
monotone properties of graphs with $\le n^{3/2-\epsilon}$ 
edges are eventually evasive.  ($n$ is the number of vertices.)

While Chowla's conjecture is not known 
to follow from the Extended Riemann Hypothesis (ERH, the
Riemann Hypothesis for Dirichlet's $L$ functions),
we show (b) with the bound $O(n^{5/4-\epsilon})$ under ERH.

We also prove unconditional results: (a$'$)
for any graph $H$, the query complexity of ``forbidden subgraph
$H$'' is $\binom{n}{2} - O(1)$; 
(b$'$) for some constant $c>0$,
all nontrivial monotone properties of graphs with 
$\le cn\log n+O(1)$ edges are eventually evasive. 

Even these weaker, unconditional results rely on deep results from
number theory such as Vinogradov's theorem on the Goldbach conjecture.  

Our technical contribution consists in connecting the topological 
framework of Kahn, Saks, and Sturtevant (1984), as further developed by
Chakrabarti, Khot, and Shi (2002), with a deeper analysis of 
the orbital structure of permutation groups and their connection to
the distribution of prime numbers.  Our unconditional results
include stronger versions and generalizations of some
result of Chakrabarti et al.
\end{abstract}

\maketitle



\footnotetext[2]{Partially supported by NSF Grant CCF-0830370.}

\section{Introduction}
\subsection{The framework}
A {\em graph property} $P_n$ of $n$-vertex graphs is 
a collection of graphs on the vertex set $[n]=\{1,\dots,n\}$ that is 
invariant under relabeling of the vertices. A property $P_n$ is called
{\em monotone} (decreasing) if it is preserved under the deletion of edges.
The trivial graph properties are the empty set and the set of all graphs.
A class of examples are the {\em forbidden subgraph} properties: for a 
fixed graph $H$, let $Q_n^H$ denote the class of $n$-vertex graphs
that do not contain a (not necessarily induced) subgraph isomorphic to $H$.


We view a set of labeled graphs on $n$ vertices as a Boolean function
on the $N=\binom{n}{2}$ variables describing adjacency.  
A Boolean function on $N$ variables is {\em evasive} if its deterministic
query (decision-tree) complexity is $N$.

The long-standing Aanderaa-Rosenberg-Karp conjecture asserts that 
{\em every nontrivial monotone graph property is evasive.}
The problem remains open even for important special classes 
of monotone properties, such as the forbidden subgraph properties.

\subsection{History}
In this note, $n$ always denotes the number of vertices of the graphs
under consideration.  

Aanderaa and Rosenberg (1973) \cite{r73} conjectured a lower bound of
$\Omega(n^2)$ on the query complexity of monotone graph
properties. Rivest and Vuillemin (1976) \cite{rv76} verified this
conjecture, proving an $n^2/16$ lower bound.  Kleitman and Kwiatkowski
(1980) \cite{kk80} improved this to $n^2/9.$ Karp conjectured that
nontrivial monotone graph properties were in fact evasive.  We refer to
this statement as the Aanderaa-Rosenberg-Karp (ARK) conjecture.

In their seminal paper, Kahn, Saks, and Sturtevant \cite{kss84}
observe that non-evasiveness of monotone Boolean functions
has strong topological consequences
(contracibility of the associated simplicial complex).
They then use results of R. Oliver about fixed points of
group actions on such complexes to verify the ARK
conjecture when $n$ is a prime-power. 
As a by-product, they improve the lower bound for general $n$ to $n^2/4.$

Since then, the topological approach of \cite{kss84}
has been influential in solving 
various interesting special cases of the ARK conjecture.
Yao (1988) \cite{y88} proves that 
non-trivial monotone properties of bipartite graphs with a given
partition $(U,V)$ are evasive (require $|U||V|$ queries).
Triesch (1996) \cite{t96} shows (in the original model) that any monotone 
property of bipartite graphs (all the graphs satisfying the property
are bipartite) is evasive.  Chakrabarti, Khot, and Shi (2002) \cite{cks02}
introduce important new techniques which we use; we improve over
several of their results (see Section~\ref{sec:main}).


\subsection{Prime numbers in arithmetic progressions} \label{sec:primes}
Dirichlet's Theorem (1837) (cf.~\cite{d80})
asserts that if $\gcd(a,m)=1$ then there exist
infinitely many primes $p\equiv a \pmod m$.  Let $p(m, a)$ denote the 
smallest such prime $p$.  
Let $p(m) =\max\{ p(m,a) \mid \gcd(a,m)=1\}$.  Linnik's celebrated
theorem (1947) asserts that $p(m) =O(m^L)$ for some absolute constant $L$
(cf.~\cite[Chap. V.]{prachar}).
Heath-Brown~\cite{heath-brown:zero-free} shows that $L\le 5.5$.  
Chowla~\cite{chowla} observes 
that under the Extended Riemann Hypothesis (ERH) we have $L \le 2+\epsilon$
for all $\epsilon > 0$ and conjectures that $L\le 1+\epsilon$
suffices: 

\begin{conjecture}[S. Chowla~\cite{chowla}] For every $\epsilon >0$
and every $m$ we have $p(m) = O(m^{1+\epsilon})$.
\end{conjecture}

This conjecture is widely believed; in fact, number theorists suggest
as plausible the stronger form $p(m) = O(m (\log m)^2)$
\cite{heath-brown:almost-primes}.  Tur\'an~\cite{turan} proves the
tantalizing result that for almost all $a$ we have $p(m,a) = O(m\log
m)$ .

Let us call a prime $p$ an {\em $\epsilon$-near Fermat prime} 
if there exists an $s\ge 0$ such that $2^s \mid p-1$ and
$\frac{p-1}{2^s} \le p^{\epsilon}$.


We need the following weak form of Chowla's
conjecture:

\begin{conjecture}[Weak Chowla Conjecture] For every $\epsilon >0$
there exist infinitely many $\epsilon$-near Fermat primes.
\end{conjecture}

In other words, the weak conjecture says that
for every $\epsilon$, for infinitely many values of $s$
we have $p(2^s, 1) < (2^s)^{1+\epsilon}$.


%

\subsection{Main results}  \label{sec:main}
For a graph property $P$ we use $P_n$ to denote the set of
graphs on vertex set $[n]$ with property $P$.  We say that 
$P$ is {\em eventually evasive} if $P_n$ is evasive for
all sufficiently large $n$.  

Our first set of results states that the ``forbidden subgraph''
property is ``almost evasive'' under three different interpretations
of this phrase.

\begin{theorem}[Forbidden subgraphs] \label{thm:forbidden}
For all graphs $H$, the forbidden subgraph property $Q_n^H$
(a) is eventually evasive, assuming the
Weak Chowla Conjecture; (b) is evasive for almost all $n$
(unconditionally); and (c)  has query complexity
$\binom{n}{2} - O(1)$ for all $n$ (unconditionally).
\end{theorem}

Part (b) says the asymptotic density of values of $n$
for which the problem is not evasive is zero.
Part (c) improves the bound $\binom{n}{2} - O(n)$ 
given in \cite{cks02}.
Parts (a) and (c) will be proved in Section~\ref{sec:forbidden}.
We defer the proof of part (b) to the journal version.

The term ``monotone property of graphs with $\le m$ edges''
describes a monotone property that fails for all graphs with more 
than $m$ edges.

\begin{theorem}[Sparse graphs]  \label{thm:density}
All nontrivial monotone properties of graphs with at most $f(n)$
edges are eventually evasive, where 
(a) under Chowla's Conjecture, $f(n)=n^{3/2-\epsilon}$
for any $\epsilon >0$; (b) under ERH, 
$f(n)=n^{5/4-\epsilon}$; and (c) unconditionally,
$f(n) = cn\log n$ for some constant $c>0$. (d)
Unconditionally, all nontrivial monotone properties of
graphs with no cycle of length greater than
$(n/4)(1-\epsilon)$ are eventually evasive
(for all $\epsilon >0$).
\end{theorem}

Part (c) of Theorem~\ref{thm:density} will be proved
in Section~\ref{sec:sparse-uncond}.  Parts (a) and (b) follow
in Section~\ref{sec:sparse-cond}.  The proof of part (d) follows
along the lines of part (c); we defer the details to
the journal version of this paper.

We note that the proofs of the unconditional results (c) and (d)
in Theorem~\ref{thm:density} rely on Haselgrove's
version~\cite{haselgrove} of Vinogradov's Theorem
on Goldbach's Conjecture (cf. Sec.~\ref{sec:vinogradov}).

Recall that
a \emph{topological subgraph} of a graph $G$ is obtained by taking
a subgraph and replacing any induced path $u - \dots -v$
in the subgraph by an edge $\{u,v\}$ (repeatedly) and deleting 
parallel edges.  A {\em minor} of a graph is obtained by 
taking a subgraph and contracting edges (repeatedly).
If a class of graphs is closed under taking minors
then it is also closed under taking topological subgraphs
but not conversely; for instance, graphs with maximum degree
$\le 3$ are closed under taking toopological subgraphs
but every graph is a minor of a regular graph of degree 3.
\begin{corollary}[Excluded topological subgraphs] \label{cor:top-sub}
Let $P$ be a nontrivial class of graphs closed under taking
topological subgraphs.  Then $P$ is eventually evasive.
\end{corollary}
This unconditional result extends one of the results of
Chakrabarti et al.~\cite{cks02}, namely, that nontrival classes of
graphs closed under taking minors is eventually evasive.

Corollary~\ref{cor:top-sub} follows from part (c) of
Theorem~\ref{thm:density} in the light of Mader's Theorem
which states that if the average degree of a graph $G$ is 
greater than $2^{\binom{k+1}{2}}$ then it contains
a topological $K_k$~\cite{mader, mader.hadw}.

Theorem~\ref{thm:density}
suggests a new stratification of the ARK
Conjecture.  For a monotone (decreasing) graph property $P_n$, let 
\[\dim(P_n) := \max\{|E(G)|-1\ |\ G \in P_n\}.\]

We can now restate the ARK Conjecture:

\begin{conjecture}
If $P_n$ is a non-evasive, non-empty, monotone decreasing graph property
then $\dim(P_n) = \binom{n}{2}-1.$
\end{conjecture}


\section{Preliminaries}

\subsection{Group action}  \label{sec:groups}
For the basics of group theory we refer to~\cite{rotman}.
All groups in this paper are finite.
For groups $\Gamma_1, \Gamma_2$ we use $\Gamma_1\le \Gamma_2$
to denote that $\Gamma_1$ is a subgroup; and 
$\Gamma_1 \lhd \Gamma_2$ to denote that $\Gamma_1$ is a
(not necessarily proper) normal subgroup.
We say that $\Gamma$ is a $p$-group if $|\Gamma|$ is a
power of the prime $p$.

For a set $\Omega$ called the ``permutation domain,'' let
$\sym(\Omega)$ denote the {\em symmetric group} on $\Omega$,
consisting of the $|\Omega|!$ permutations of $\Omega$.
For $\Omega=[n]=\{1,\dots,n\}$, we set $\Sigma_n = \sym([n])$.
For a group $\Gamma$, a homomorphism $\vf\,:\,\Gamma\to\sym(\Omega)$
is called a {\em $\Gamma$-action} on $\Omega$.  The action is {\em faithful} 
if $\ker(\vf)=\{1\}$.  For $x\in\Omega$ and $\gamma\in\Gamma$ we denote by 
$x^{\gamma}$ the image of $x$ under $\vf(\gamma)$.  For $x\in\Omega$ we write 
$x^{\Gamma} = \{ x^{\gamma}\,:\,\gamma\in\Gamma\}$ and call it the 
{\em orbit} of $x$ under the $\Gamma$-action.  The orbits partition $\Omega$.

Let $\binom{\Omega}{t}$ denote the set of $t$-subsets of $\Omega$.  There
is a natural induced action $\sym(\Omega)\to\sym(\binom{\Omega}{t})$ which
also defines a natural $\Gamma$-action on $\binom{\Omega}{t}$.
We denote this action by $\Gamma^{(t)}$.
Similarly, there is a natural induced $\Gamma$-action on $\Omega\times\Omega$.
The orbits of this action are called the {\em orbitals} of $\Gamma$.
We shall need the undirected version of this concept; we shall call
the orbits of the $\Gamma$-action on $\binom{\Omega}{2}$ the {\em u-orbitals}
(undirected orbitals) of the $\Gamma$-action.

By an action of the group $\Gamma$ on a structure $\mathfrak X$ such as a
group or a graph or a simplicial complex we mean a homomorphism
$\Gamma\to \aut({\mathfrak X})$ where $\aut({\mathfrak X})$ denotes the
automorphism group of $\mathfrak X$.

Let $\Gamma$ and $\Delta$ be groups and let 
$\psi \, : \, \Delta \to \aut(\Gamma)$ 
be a $\Delta$-action on $\Gamma$.  These data uniquely define a group 
$\Theta = \Gamma\rtimes \Delta$,
the \emph{semidirect product} of $\Gamma$ and $\Delta$ with respect to $\psi$.
This group has 
order $|\Theta|=|\Gamma||\Delta|$ and has the following properites: $\Theta$ has 
two subgroups $\Gamma^*\cong \Gamma$ and 
$\Delta^*\cong \Delta$ such that $\Gamma^*\lhd \Theta$;\
$\Gamma^*\cap \Delta^* =\{1\}$; 
and $\Theta=\Gamma^*\Delta^*=\{\gamma\delta \mid \gamma\in \Gamma^*,
\delta\in \Delta^*\}$.
Moreover, identifying $\Gamma$ with $\Gamma^*$ and $\Delta$ with $\Delta^*$,
for all $\gamma\in \Gamma$ and $\delta\in \Delta$ we have 
$\gamma^{\psi(\delta)}= \delta^{-1}\gamma\delta$.

$\Theta$ can be defined as the set 
$\Delta\times \Gamma$ under the group operation
$$ (\delta_1,\gamma_1)(\delta_2,\gamma_2) =
 (\delta_1\delta_2, \gamma_1^{\psi(\delta_2)}\gamma_2) \quad\quad
(\delta_i\in \Delta, \gamma_i\in \Gamma).$$

For more on semidirect products, which we use extensively,
see~\cite[Chap. 7]{rotman}.

The group $\agl(1,q)$ of affine transformations $x\mapsto ax+b$
of $\F_q$ ($a\in \F_q^{\times}$, $b\in \F_q$) acts on $\F_q$.
For each $d\mid q-1$,\ $\agl(1,q)$ has a unique subgroup of order
$qd$; we call this subgroup $\Gamma(q,d)$.  We note that
$\F_q^+\lhd \Gamma(q,d)$ and $\Gamma(q,d)/\F_q^+$ is cyclic of order $d$
and is isomorphic to a subgroup $\Delta$ of $\agl(1,q)$;
$\Gamma(q,d)$ can be described as a {\em semidirect product}
$(\F_q^+) \rtimes \Delta$.

\subsection{Simplicial complexes and monotone graph properties}

\noindent 
An {\em abstract simplicial complex} $\calK$ on the set $\Omega$ 
is a subset of the power-set of $\Omega$, closed under subsets:
if $B\subset A\in{\calK}$ then $B\in{\calK}$.  
The elements of $\calK$
are called its {\em faces}.  The {\em dimension} of $A\in{\calK}$
is $\dim(A) = |A|-1$; the dimension of $\calK$ is 
$\dim({\calK})=\max \{\dim(A)\mid A\in{\calK}\}$.  The
{\em Euler characteristic} of ${\calK}$ is defined as
\[ \chi({\calK}) := \sum_{A \in {\calK}, A \ne \emptyset} {(-1)^{\dim(A)}}.\]
Let $[n] := \{1,2, \ldots, n\}$ and $\Omega=\binom{[n]}{2}$.  Let $P_n$ be
a subset of the power-set of $\Omega$, \ie, a set of graphs on the
vertex set $[n]$.  We call $P_n$ a {\em graph property} if it is invariant 
under the induced action $\Sigma_n^{(2)}$.
We call this graph property {\em monotone decreasing} if it is
closed under subgraphs, \ie, it is a simplicial complex.
We shall omit the adjective ``decreasing.''

\subsection{\bf Oliver's Fixed Point Theorem}
Let $\calK \subseteq 2^{\Omega}$ be an abstract simplicial complex
with a $\Gamma$-action.   The {\em fixed point complex} $\calK_{\Gamma}$
action is defined as follows.
Let $\Omega_1, \dots, \Omega_k$ be the $\Gamma$-orbits on $\Omega$. Set
\[\calK_{\Gamma} := \{S \subseteq [k] \mid 
\bigcup_{i \in S} \Omega_i \in \calK \}.\]

\noindent
We say that a group $\Gamma$ satisfies {\bf Oliver's condition}
if there exist (not necessarily distinct) primes $p,q$ such that
$\Gamma$ has a (not necessarily proper)
chain of subgroups $\Gamma_2 \lhd \Gamma_1 \lhd \Gamma$
such that $\Gamma_2$ is a $p$-group, $\Gamma_1/\Gamma_2$ is cyclic,
and $\Gamma/\Gamma_1$ is a $q$-group.

\begin{theorem}[Oliver~\cite{o75}]  \label{thm:oliver}
Assume the group $\Gamma$ satisfies Oliver's condition.
If $\Gamma$ acts on a nonempty contractible simplicial 
complex $\calK$ then 
\begin{equation}
\chi(\calK_{\Gamma}) \equiv 1 \pmod q.
\end{equation}
\end{theorem}
In particular, such an action must always have a nonempty invariant
face.

\subsection{The KSS approach and the general strategy}
The topological approach to evasiveness, initiated by 
Kahn, Saks, and Sturtevant, is based on the following
key observation.
\begin{lemma}[Kahn-Saks-Sturtevant~\cite{kss84}] \label{lem:kss}
If $P_n$ is a non-evasive graph property then $P_n$ 
is contractible.
\end{lemma}
Kahn, Saks, and Sturtevant recognized that Lemma~\ref{lem:kss}
brought Oliver's Theorem to bear on evasiveness.
The combination of Lemma~\ref{lem:kss} and Theorem~\ref{thm:oliver}
suggests the following general strategy, used by all authors
in the area who have employed the topological method, including
this paper: We find primes $p,q$, 
a group $\Gamma$ satisfying Oliver's condition with these primes,
and a $\Gamma$-action on $P_n$, such that $\chi(P_n)\equiv 0\pmod q$.
By Oliver's Theorem and the KSS Lemma this implies that $P_n$ is
evasive.  The novelty is in finding the right $\Gamma$.

KSS~\cite{kss84} made the assumption that $n$ is a prime power
and used as $\Gamma = \agl(1,n)$, the group of affine transformations
$x\mapsto ax+b$ over the field of order $n$.  While we use 
subgroups of such groups as our building blocks, the attempt
to combine these leads to hard problems on the distribution of
prime numbers.  

Regarding the ``forbidden subgraph'' property, Chakrabarti, Khot, 
and Shi~\cite{cks02} built considerable machinery which we use.
Our conclusions are considerably stronger than theirs;
the additional techniques involved include a study of the
orbitals of certain metacyclic groups, a universality property
of cyclotomic graphs derivable using Weil's character sum estimates,
plus the number theoretic reductions indicated.

For the ``sparse graphs'' result (Theorem~\ref{thm:density}) we need
$\Gamma$ such that all u-orbitals of $\Gamma$ are large and therefore
$(P_n)_{\Gamma}=\{\emptyset\}$.  

In both cases, we are forced to use rather large building blocks
of size $q$, say, where $q$ is a prime such that $q-1$ has a large
divisor which is a prime for Theorem~\ref{thm:density} and a power
of 2 for Theorem~\ref{thm:forbidden}.

\section{Forbidden subgraphs}  \label{sec:forbidden}
In this section we prove parts (a) and (c) of
Theorem~\ref{thm:forbidden}.

\subsection{The CKS condition}
A {\em homomorphism} of a graph $H$ to a graph $H'$ is a map
$f\,:\,V(H)\to V(H')$ such that 
$(\forall x,y\in V(H))(\{x,y\}\in E(H) \Rightarrow 
\{f(x),f(y)\}\in E(H'))$.  (In particular, $f^{-1}(x')$ is an 
independent set in $H$ for all $x'\in V(H')$.)
Let $Q_r^{[[H]]}$ be the set of those $H'$ with $V(H')=[r]$
that do not admit an $H\to H'$ homomorphism.
Let further $T_H := \min\{2^{2^t} - 1\ \mid \ 2^{2^t} \geq h\}$
where $h$ denotes the number of vertices of $H$.
The following is the main lemma of Chakrabarti, Khot, and Shi~\cite{cks02}.
\begin{lemma}[Chakrabarti et al.~\cite{cks02}] \label{lem:even_char}
If $r \equiv 1 \pmod{T_H}$ then $\chi(Q_r^{[[H]]}) \equiv 0 \pmod 2$.
\end{lemma}

\subsection{Cliques in generalized Paley graphs}
Let $q$ be an odd prime power and
$d$ an even divisor of $q-1.$ 
Consider the graph $P(q,d)$ whose vertex set is $\mathbb{F}_q$
and the adjacency between the vertices is defined as follows: 
$i \sim j \iff (i-j)^d = 1.$
$P(q,d)$ is called a {\em generalized Paley graph}.
\begin{lemma}   \label{lem:paley_clique}
If $(q-1)/d \leq q^{1/(2h)}$ then 
$P(q, d)$ contains a clique on $h$ vertices. 
\end{lemma}
This follows from the following lemma which in turn can be proved by
a routine application of Weil's character sum estimates (cf.~\cite{bgw}).
\begin{lemma}   \label{extension_lemma}
Let $a_1, \ldots, a_t$ be distinct elements of the finite field
$\mathbb{F}_q.$  Assume $\ell \mid q-1$.
Then the number of solutions 
$x \in \mathbb{F}_q$ to the system of equations $(a_i + x)^{(q-1)/\ell} =1$
is $\frac{q}{\ell^t} \pm t \sqrt q.$  \qed
\end{lemma}

%
Let $\Gamma(q,d)$ be the subgroup of order $qd$ of $\agl(1,q)$
defined in Section~\ref{sec:groups}.
\begin{observation}
Each u-orbital of $\Gamma(q,d)$ is isomorphic to $P(q,d)$. \qed
\end{observation}
\begin{corollary} 
If ${\frac{q-1}{d} \leq q^{1/(2h)}}$ then 
each u-orbital of $\Gamma(q,d)$ contains a clique of size $h.$ 
\end{corollary}

\subsection{ $\epsilon$-near-Fermat primes}
The numbers in the title were defined in Section~\ref{sec:primes}.
In this section we prove Theorem~\ref{thm:forbidden}, part (a).


\begin{theorem} Let $H$ be a graph 
on $h$ vertices. If there are infinitely many   
$\frac{1}{2h}$-near-Fermat primes then 
$Q_n^H$ is eventually evasive.
\label{near-fermat}
\end{theorem}
\noindent{\em Proof.} 
Fix an odd prime $p \equiv 2 \pmod{T_H}$ such that $p \geq |H|.$ 
If there are infinitely many $\frac{1}{2h}$-near-Fermat primes then
infinitely many of them belong to the same residue class mod~$p$,
say $a+\Z p$.  
Let $q_i$ be the $i$-{th} $\frac{1}{2h}$-near-Fermat prime such that
$q_i \geq p$ and $q_i \equiv a \pmod p.$ 
Let $r' = na^{-1} \pmod p$ and $k' = \sum_{i=1}^{r'} q_i.$
Then $k'\equiv n \pmod p$ and therefore
$n = pk + k'$ for some $k$.

Now in order to use Lemma~\ref{lem:even_char}, we need to write
$n$ as a sum of $r$ terms where $r\equiv 1 \pmod{T_H}$.  We already have
$r'$ of these terms; we shall choose each of the remaining $r-r'$ terms
to be $p$ or $p^2$.  If there are $t$ terms equal to $p^2$ then this
gives us a total of $r=t+(k-tp)+r'$ terms, so we need 
$t(p-1)\equiv k+r' \pmod{T_H}$.  By assumption, $p-1\equiv 1\pmod{T_H}$;
therefore such a $t$ exists; for large enough $n$, it will also satisfy
the constraints $0\le t\le k/p$,

Let now
\[\Lambda_1 := \left((\mathbb{F}^+_{p^2})^{t} \times 
(\mathbb{F}^+_p)^{k - tp}\right) \rtimes \mathbb{F}_{p^2}^{\times}\]
acting on $[pk]$ with $t$ orbits of size $p^2$ and $k-pt$ orbits
of size $p$ as follows: on an orbit of size $p^i$ ($i=1,2$) the
action is $\agl(1,p^i)$.  The additive groups act independently,
with a single multiplicative action on top.  $\F_{p^2}^{\times}$
acts on $\F_p^+$ through the group homomorphism 
$\F_{p^2}^{\times}\to \F_{p}^{\times}$ defined by the map $x\mapsto x^{p-1}$.
Let $B_j$ denote an orbit of $\Lambda_1$ on $[kp]$.
Now the orbit of any pair $\{u,v\} \in {B_j \choose 2}$
is a clique of size $|B_j|\ge p \ge h$, therefore 
a $\Lambda_1$-invariant graph cannot contain an
intra-cluster edge.

Let $d_i$ be the largest power of 2 that divides $q_i-1.$
Let $C_i$ be the subgroup of $\mathbb{F}_{q_i}^\times$ of order $d_i.$
Let
$\displaystyle{
\Lambda_2 := \prod_{i=1}^{r'} \Gamma(q_i,d_i),
}$
acting on $[k']$ with $r'$ orbits of sizes $q_1,\dots,q_{r'}$
in the obvious manner.

From Lemma~\ref{lem:paley_clique}
we know that the orbit of any $\{u,v\} \in {[q_i] \choose 2}$
must contain a clique of size $h.$ 
Hence, an invariant graph cannot contain
any intra-cluster edge.

Overall, let $\Gamma := \Lambda_1 \times \Lambda_2$,
acting on $[n].$
Since $q_i \geq p,$ we have $\gcd(q_i, p^2-1) = 1.$
Thus, $\Gamma$ is a ``$2$-group extension of a cyclic extension of
a $p$-group'' and therefore satisfies Oliver's Condition
(stated before Theorem~\ref{thm:oliver}). Hence, assuming $Q_n^H$ is
non-evasive, Lemma~\ref{lem:kss} and Theorem~\ref{thm:oliver}
imply
\[\chi((Q_n^H)_{\Gamma}) \equiv 1 \pmod 2.\]

On the other hand, we claim that the fixed-point complex
$(Q_n^H)_{\Gamma}$ is isomorphic to $Q_{r}^{[[H]]}$.
The (simple) proof goes along the lines of 
Lemma~4.2 of \cite{cks02}. Thus, by Lemma~\ref{lem:even_char}
we have $\chi(Q_r^{[[H]]}) \equiv 0 \pmod 2,$  a contradiction. \qed

\subsection{Unconditionally, $Q_n^H$ is only $O(1)$ away from being evasive}

\noindent
In this section, we prove part (c) of Theorem~\ref{thm:forbidden}.
\begin{theorem} \label{near_eva}
For every graph $H$ there exists a number $C_H$ such that the
query complexity of  $Q_n^H$ is $\ge \binom{n}{2} - C_H.$
\end{theorem}
\noindent {\em Proof.}
Let $h$ be the number of vertices of $H$.
Let $p$ be the smallest prime such that $p \geq h$ and 
$p \equiv 2 \pmod{T_H}$.  So $p < f(H)$ for some function $f$
by Dirichlet's Theorem (we don't need any specific estimates here).
Since $p-1 \equiv 1 \pmod{T_H},$ we have $\gcd(p-1, T_H) = 1$
and therefore $\gcd(p-1, pT_H) = 1$.  Now, by the Chinese
Remainder Theorem, select the smallest positive integer $k'$
satisfying $k'\equiv n \pmod{pT_H}$ and $k'\equiv 1 \pmod{p-1}$.
Note that $k' < p^2T_H$.   Let $k = (n-k')/(pT_H)$; so we have
$n = kp T_H  + k'$.

Let $N' = \binom{n}{2} - \binom{k'}{2}$.  
Consider the following Boolean function $B_n^H$ on $N'$
variables.   Consider graphs $X$ on the vertex set $[n]$
with the property that they have no edges among their
last $k'$ vertices.   These graphs can be viewed as
Boolean functions of the remaining $N'$ variables.
Now we say that such a graph has property $B_n^H$
if it does not contain $H$ as a subgraph.

\noindent
{\bf Claim.} The function $B_n^H$ is evasive.

\noindent
The Claim immediately implies that the query complexity
of $Q_n^H$ is at least $N'$, proving the Theorem
with $C_H = \binom{k'}{2} < p^4 T_H^2 < f(H)^4 T_H^2$.

To prove the Claim, consider the groups
$\Lambda := (\mathbb{F}_p^+)^{kT_H} \rtimes \mathbb{F}_p^{\times}$
and $\Gamma := \Lambda \times \mathbb{Z}_{k'}$.
Here $\Lambda$ acts on $[pkT_H]$ in the obvious way:
we divide $[pkT_H]$ into $kT_H$ blocks of size $p$;
$\mathbb{F}_p^+$ acts on each block independently
and $\mathbb{F}_p^{\times}$ acts on the blocks simultaneously (diagonal
action) so on each block they combine to an $\agl(1,p)$-action.
$\mathbb{Z}_{k'}$ acts as a $k'$-cycle on the remaining $k'$
vertices.
So $\Gamma$ is a cyclic extension of a $p$-group (because
$\gcd(p-1,k')=1$).

If $B_n^H$ is not evasive then 
from Theorem~\ref{thm:oliver} and Lemma~\ref{lem:kss},
we have  $\chi\left((B_n^H)_{\Gamma}\right) = 1$.

On the other hand we claim that,
$(B_n^H)_{\Gamma} \cong Q_{r}^{[[H]]},$
where $r = k T_H + 1.$
The proof of this claim is exactly the same as the proof
of Lemma~4.2 of \cite{cks02}.  Thus, from Lemma~\ref{lem:even_char},
we conclude that
$\chi(Q_r^{[[H]]})$
 is even. This contradicts the previous conclusion that
$\chi(Q_r^{[[H]]}) = 1.$  \qed

\begin{remark}
Specific estimates on the smallest Dirichlet prime can be used
to estimate $C_H$.  Linnik's theorem implies $C_H < h^{O(1)}$,
extending Theorem~\ref{near_eva} to strong lower bounds for
variable $H$ up to $h = n^c$ for some positive constant $c$.
\end{remark}

\section{Sparse graphs: unconditional results} \label{sec:sparse-uncond}
We prove part (c) of Theorem~\ref{thm:density}.
\begin{theorem}
If the non-empty monotone graph property $P_n$ is not evasive
then 
$$\dim(P_n) = \Omega(n \log n).$$
\label {superlin}
\end{theorem}
\subsection{The basic group construction}\label{wreath-with-autos}
Assume in this section that
$n = p^{\alpha} k$ where $p$ is prime.  Let $\Delta_k\le \Sigma_k$.
We construct the group $\Gamma_0(p^{\alpha}, \Delta_k)$ acting on $[n].$

Let $\Delta = (\mathbb{F}_{p^{\alpha}}^\times \times \Delta_k)$.
Let $\Gamma_0(p^{\alpha}, \Delta_k)$ be the semidirect product 
$(\F_{p^{\alpha}}^+)^k \rtimes \Delta$
with respect to the $\Delta$-action on $(\F_{p^\alpha}^+)^k$ defined by
\[(a, \sigma) : (b_1, \ldots, b_k) \mapsto (ab_{\sigma^{-1}(1)},
\ldots, ab_{\sigma^{-1}(k)}).\]


We describe the action of $\Gamma_0(p^{\alpha}, \Delta_k)$ on $[n]$.
Partition $[n]$ into $k$ clusters of size $p^{\alpha}$ each.
Identify each cluster with the field of order $p^{\alpha},$
i.e., as a set, $[n] = [k] \times \mathbb{F}_{p^{\alpha}}.$
The action of $\gamma = (b_1, \ldots, b_k, a, \sigma)$ is described by
\[\gamma : (x, y) \mapsto (\sigma(x), ay + b_{\sigma(x)}).\]

An unordered pair $(i,j) \in [n]$ is termed an {\em
intra-cluster edge} if both $i$ and $j$ are in the same
cluster, otherwise it is termed an {\em inter-cluster
edge.} Note that every u-orbital under $\Gamma$ has
only intra-cluster edges or only inter-cluster
edges. Denote by $m_{\operatorname{intra}}$ and
$m_{\operatorname{inter}}$ the minimum sizes of
u-orbitals of intra-cluster and inter-cluster edges
respectively.

We denote by $m'_k$ the minimum size of an orbit in $[k]$ under $\Delta_k$ and 
by $m''_k$ the minimum size of a u-orbital in $[k].$ We then have:

\[ m_{\operatorname{intra}} \geq \binom{p^\alpha}{2} \times m'_k, \qquad m_{\operatorname{inter}} \geq (p^\alpha)^2 \times m''_k \]

Let $m_k `:= \min \{m'_k, m''_k \}$ and define $m^*$ 
as the minimum size of a u-orbital in $[n].$ Then

\begin{equation}  \label{eq:mstarequation} 
 m^* = \min \{ m_{\operatorname{intra}}, m_{\operatorname{inter}} \} 
= \Omega(p^{2\alpha} m_k) 
\end{equation}

\subsection{Vinogradov's Theorem}   \label{sec:vinogradov}
The Goldbach Conjecture asserts that every even integer
can be written as the sum of two primes.  Vinogradov's
Theorem \cite{v37} says that every sufficiently large
odd integer $k$ is the sum of three primes $k=p_1+p_2+p_3$.
We use here Haselgrove's version~\cite{haselgrove} of Vinogradov's theorem 
which states that we can require the primes to be roughly equal:
$p_i\sim k/3$.  This can be combined with the Prime Number Theorem
to conclude that every sufficiently
large even integer $k$ is a sum of four roughly equal primes.


\subsection{Construction of the group}
\label{semidirect product with additive partition}
Let $n = p^\alpha k$ where $p$ is prime.   Assume $k$ is
not bounded.
Write $k$ as a sum of $t\le 4$ roughly equal primes $p_i$.
Let $\Delta_k := \prod_i \mathbb{Z}_{p_i} $
where $\mathbb{Z}_{p_i}$ denotes the cyclic group of order $p_i$
and the direct product is taken over the {\em distinct} $p_i$.

$\Delta_k$ acts on $[k]$ as follows: partition $k$ into
parts of sizes $p_1,\dots,p_t$ and call these parts $[p_i].$
The group $\mathbb{Z}_{p_i}$ acts as a cyclic group on the part
$[p_i].$  In case of repetitions, the same
factor $\mathbb{Z}_{p_i}$ acts on all the parts of size $p_i.$

We follow the notation of Section~\ref{wreath-with-autos}
and consider the group $\Gamma_0(p^{\alpha}, \Delta_k)$ with
our specific $\Delta_k$.  We have $m_k = \Omega(k)$ and
hence we get, from equation~\eqref{eq:mstarequation}:
\begin{lemma}\label{additive partition minsize lemma}
Let $n = p^\alpha k$ where $p$ is a prime. For the
group $\Gamma_0(p^\alpha,\Delta_k)$, we have 
$m^* = \Omega(p^{2\alpha}k) = \Omega(p^\alpha n),$ 
where $m^*$ denotes the minimum size of a u-orbital.
\end{lemma}

\subsection{Proof for the superlinear bound} \label{sec:super}
Let $n = p^{\alpha} k$ where $p^{\alpha}$ is
the largest prime power dividing $n$;
so $p^{\alpha} = \Omega(\log n)$; this will be
a lower bound on the size of u-orbitals.
Our group $\Gamma$ will be 
of the general form discussed in Section~\ref{wreath-with-autos}.

\noindent
{\sf Case 1.} $p^{\alpha} = \Omega(n^{2/3}).$ \\
Let $\Gamma = \Gamma_0(p^{\alpha}, \{1\})$.
Following the notation of Section~\ref{wreath-with-autos}, 
we get $m_k' = m_k'' = 1,$ and this yields that 
$m^* = \Omega((p^{\alpha})^2) = \Omega(n^{4/3}) = \Omega(n \log n).$
Oliver's condition is easily verified for $\Gamma$.
%

\noindent
{\sf Case 2.} $k = \Omega (n^{1/3}).$ \\
Consider the $\Gamma:=\Gamma_0(p^\alpha,\Delta_k)$ acting on
$[n]$ where $\Delta_k$ is as described in
Section~\ref{semidirect product with additive partition}. 
The minimum possible size $m^*$ of
a u-orbital is $\Omega(np^\alpha)$ by 
Lemma~\ref{additive partition minsize lemma}.
Finally, since $p^\alpha = \Omega(\log n)$, we obtain
$m^* = \Omega(n\log n).$

If all $p_i$ are co-prime to $p^{\alpha} -1$ 
then $\mathbb{F}_{p^{\alpha}}^{\times} \times \Delta_k$
becomes a cyclic group and $\Gamma$ becomes a cyclic
extension of a $p$-group.

Since $p_i = \Omega(k) = \Omega(n^{1/3})$ for all $i$
and $p^\alpha = O(n^{2/3})$, size considerations yield
that at most one $p_i$ divides $p^{\alpha} - 1$ and
$p_i^2$ does not. Suppose, without loss of generality,
$p_1$ divides $p^{\alpha} - 1.$ Let $p^{\alpha} - 1 =
p_1 d,$ then $d$ must be co-prime to each $p_i.$ Thus,
$\Delta = (\Z_{p_1} \times \Z_d) \times (\Z_{p_1} \times
\ldots \times \Z_{p_t}) = (\Z_{d} \times \Z_{p_2}
\times \ldots \times \Z_{p_r}) \times (\Z_{p_1} \times
\Z_{p_1}).$ Thus, $\Delta$ is a $p_1$-group extension of a
cyclic group.  Hence, $\Gamma$ satisfies Oliver's
Condition (cf. Theorem~\ref{thm:oliver}).  
\qed


%
%
\begin{remark}
For almost all $n,$ our proof gives a better dimension lower bound of 
$\Omega(n^{1 + \frac{1+o(1)}{\ln \ln n}}).$
\end{remark}

\section{Sparse graphs: conditional improvements}  \label{sec:sparse-cond}
In this section we prove parts (a) and (b) of Theorem~\ref{thm:density}.

\subsection{General Setup}\label{sparse-conditional setup}

Let $n = pk + r,$ where $p$ and $r$ are prime
numbers. Let $q$ be a prime divisor of $(r-1).$ We
partition $[n]$ into two parts of size $pk$ and $r$,
denoted by $[pk]$ and $[r]$ respectively. We now
construct a group $\Gamma(p,q,r)$ acting on $[n]$ as a
direct product of a group acting on $[pk]$ and a group
acting on $[r],$ as follows:
\[ \Gamma = \Gamma(p,q,r) := \Gamma_0(p,\Delta_k) \times \Gamma(r,q) \]
Here, $\Gamma_0(p,\Delta_k)$ acts on $[pk]$ and is as defined
in Section~\ref{semidirect product with additive
partition}, and involves choosing a partition of $k$
into upto four primes that are all $\Omega(k).$

$\Gamma(r,q)$ is defined as the semidirect product
$\mathbb{F}_r^+ \rtimes C_q,$ with $C_q$ viewed as a
subgroup of the group $\mathbb{F}_r^\times.$ It acts on
$[r]$ as follows: We identify $[r]$ with the field of
size $r.$ Let $(b,a)$ be a typical element of
$\Gamma_r$ where $b \in \mathbb{F}_r$ and $a \in C_q.$
Then, $(b,a):x \mapsto ax + b.$

Thus, $\Gamma = \Gamma(p,q,r)$ acts on $[n].$ Let $m^*$
be the minimum size of the orbit of any edge $(i,j) \in
{[n] \choose 2}$ under the action of $\Gamma.$ One can
show that
\begin{equation}
 m^* = \Omega(\min\{p^2k, pkr, qr\}).
\label{pqr_bound}
\end{equation}
We shall choose $p,q,r$ carefully such that
(a) the value of $m^*$ is large, and
(b) Oliver's condition holds for $\Gamma(p,q,r)$.
\subsection{ERH and Dirichlet primes}
\label{grh}
The Extended Riemann Hypothesis (ERH) implies the
following strong version of the Prime Number Theorem
for arithmetic progressions.  Let $\pi(n,D,a)$ denote 
the numer of primes $p\le n$, $p\equiv a \pmod D$.
Then for $D < n$ we have
\begin{equation}
   \pi(n,D,a) = \frac{\li(n)}{\varphi(D)} + O(\sqrt{x}\ln x)
\end{equation}
where $\li(n)=\int_2^n dt/t$ and the constant implied by the
big-Oh notation is absolute (cf. \cite[Ch.\,7, eqn.\,(5.12)]{prachar}
or \cite[Thm.\,8.4.5]{bach}).

This result immediately implies ``Bertrand's Postulate for
Dirichlet primes:''
\begin{lemma}[Bertrand's Postulate for 
Dirichlet primes]\label{dirichlet-bertrand}
Assume ERH.  Suppose the sequence $D_n$ satisfies 
$D_n = o(\sqrt{n}/\log^2 n)$.  Then for all sufficiently
large $n$ and for any $a_n$ relatively prime to $D_n$
there exists a prime $p \equiv a_n \pmod{D_n}$ such that
$\frac{n}{2} \leq p \leq n.$ 
\end{lemma}
\subsection{With ERH but without Chowla}
We want to write $n = pk + r,$ where $p$ and $r$ are
primes, and with $q$ a prime divisor of $r - 1,$ as
described in Section~\ref{sparse-conditional
setup}. Specifically, we try for:
\[ p = \Theta(n^{1/4}),\quad
 \frac{n}{4} \leq r \leq \frac{n}{2}, \quad
q = \Theta(n^{1/4 - \epsilon})\]
We claim that under ERH, such a partition of $n$ is
possible.

To see this, fix some $p = \Theta(n^{1/4})$
such that $\gcd(p, n) = 1.$ Fix some $q =
\Theta(n^{1/4 - \epsilon}).$ Now,
$r \equiv 1 \pmod q$ and $r \equiv n \pmod p$ solves to 
$r \equiv a \pmod{pq}$ for some $a$ such that $\gcd(a, pq) = 1.$ 
Since $pq = \Theta(n^{1/2 - \epsilon}),$ 
we can conclude under ERH (using Lemma~\ref{dirichlet-bertrand})
that there exists a prime 
$r \equiv a \pmod{pq}$ such that $\frac{n}{4} \leq r \leq
\frac{n}{2}.$ This gives us the desired partition.
%
One can verify that our $\Gamma$ 
satisfies Oliver's Condition. 
Equation~\eqref{pqr_bound} gives 
$m^* = \Omega(n^{5/4-\epsilon}).$ 
This completes the proof of part (b) of Theorem~\ref{thm:density}. \qed

\subsection{Stronger bound using Chowla's conjecture}
Let $a$ and $D$ be relatively prime.
Let $p$ be the first prime such that $p \equiv a \pmod D.$ 
Chowla's conjecture tells us that $p = O(D^{1+\epsilon})$ for 
every $\epsilon > 0.$ Using this, we show
$m^* = \Omega(n^{3/2-\epsilon}).$

We can use Chowla's conjecture, along with the general 
setup of Section~\ref{sparse-conditional setup}, 
to obtain a stronger lower bound on $m^*.$ 
The new bounds we hope to achieve are:
\[ p = \Theta(\sqrt{n}),\quad n^{1-2.5\delta} \leq r \leq n^{1- 0.5\delta},
\quad q = \Theta(n^{1/2 - \delta}) \]
Such a
partition is always possible assuming Chowla's conjecture.
To see this, first fix $p = \Theta(n^{1/2}),$ then
fix $q = \Theta(n^{1/2 - 2\delta})$ and find the least solution 
for $r \equiv 1 \pmod q$ and $r \equiv n \pmod p,$
which is equivalent to solving for $r \equiv a \pmod {pq}$ for 
some $a < pq.$ The least solution will be greater than $pq$ unless
$a$ happens to be a prime. In this case, we add another constraint,
say $r \equiv a+1 \pmod 3$ and resolve to get the least solution
greater than $pq.$ Note that $n^{1-2.5\delta} \leq r \leq n^{1-0.5\delta}.$
Now, from Equation (\ref{pqr_bound}), we get 
the lower bound of $m^* = \Omega(n^{3/2 - 4 \delta}).$ 
This completes the proof of part (a) of Theorem~\ref{thm:density}. \qed


\subsection*{Acknowledgment.}  Raghav Kulkarni expresses his
gratitude to Sasha Razborov for bringing the subject to his attention
and for helpful initial discussions.


\begin{thebibliography}{99}
	
\bibitem {bgw} Babai, L., G\'al, A., Wigderson, A.:
   Superpolynomial lower bounds for monotone span programs.
   {\em Combinatorica} 19 (1999), 301--320.


\bibitem {bach} Bach, E., Shallit, J.:
{\em Algorithmic Number Theory, Vol. 1.}
The MIT Press 1996.




\bibitem {cks02}{Chakrabarti, A., Khot, S., Shi, Y.:
{Evasiveness of Subgraph Containment and Related Properties.}
{\em SIAM J. Comput.} 31(3) (2001), 866-875.}

\bibitem {chowla}{Chowla, S.} 
On the least prime in the arithmetical progression.
{\em J. Indian Math. Soc.} 1(2) (1934), 1--3.

\bibitem{d80}{Davenport, H.:
{\em Multiplicative Number Theory.} (2nd Edn) 
Springer Verlag, New York, 1980.}


\bibitem{gp90} {Granville, A., Pomerance, C.:
On the least prime in certain arithmetic progressions.
{\em J. London Math. Soc.} 41(2) (1990), 193--200.}

\bibitem {haselgrove} Haselgrove, C. B.:
Some theorems on the analytic theory of numbers.
\emph{J. London Math. Soc.} 36 (1951) 273--277

\bibitem {heath-brown:almost-primes} Heath-Brown, D. R.:
{Almost-primes in arithmetic progressions and short intervals.}
{\em Math. Proc. Cambr. Phil. Soc.} 83 (1978) 357--376.

\bibitem {heath-brown:zero-free} Heath-Brown, D. R.:
Zero-free regions for Dirichlet $L$-functions, and the least
prime in an arithmetic progression.
{\em Proc. London Math. Soc.} 64(3) (1992) 265--338.

\bibitem {kk80} {Kleitman, D. J., Kwiatkowski, D. J.:
{Further results on the Aanderaa-Rosenberg Conjecture}
{\em J. Comb. Th. B} 28 (1980), 85--90.}

\bibitem {kss84}{Kahn, J., Saks, M., Sturtevant, D.:
{A topological approach to evasiveness.} 
{\em Combinatorica} 4 (1984), 297--306.}




\bibitem {l02}{Lutz, F. H.: 
{Examples of $\mathbb{Z}$-acyclic and contractible vertex-homogeneous
simplicial complexes.}. 
{\em Discrete Comput. Geom.} 27 (2002), No. 1, 137--154.}

\bibitem {mader}  Mader, W.:
  Homomorphieeigenschaften und mittlere Kantendichte von Graphen.
 {\em Math. Ann.} 174 (1967), 265--268.

 \bibitem {mader.hadw} Mader, W.:
  Homomorphies\"atze f\"ur Graphen.
 {\em Math. Ann.} {\bf 175} (1968), 154--168.

\bibitem {o75}{Oliver, R.: {Fixed-point sets of group actions on 
finite acyclic complexes}. {\em  Comment. Math. Helv.}
50 (1975), 155--177.}

\bibitem {prachar} Prachar, K.:
 {\em Primzahlverteilung.}  
 Springer, 1957.

\bibitem {r73} {Rosenberg A. L.:} 
{On the time required to recognize properties of graphs: A problem.}
{\em SIGACT News} 5 (4) (1973), 15--16.

\bibitem {rotman} Rotman, J.: 
{\em  An Introduction to the Theory of Groups.} Springer Verlag, 1994.

\bibitem {rv76}{Rivest, R.L., Vuillemin, J.:}
{On recognizing graph properties from adjacency matrices}.
{\em Theoret. Comp. Sci.} 3 (1976), 371--384.

\bibitem {s41} {Smith P. A.: 
{Fixed point theorems for periodic transformations}.
{\em Amer. J. of Math.} 63 (1941), 1--8.}

\bibitem {titchmarsh} Titchmarsh, E. C.:
A divisor problem.
{\em Rend. Circ. Mat. Palermo} 54 (1930), 419--429.

\bibitem {t96}{Triesch, E.: 
{On the recognition complexity of some graph properties}.
{\em Combinatorica} 16 (2) (1996) 259--268.}

\bibitem {turan} Tur\'an, P.:
\"Uber die Primzahlen der arithmetischen Progression.
{\em Acta Sci. Math. (Szeged)} 8 (1936/37) 226--235.


\bibitem {v37} {Vinogradov, I. M.:
{\em The Method of Trigonometrical Sums in the Theory of Numbers (Russian).}
Trav. Inst. Math. Stekloff 10, 1937.}

\bibitem {y88} {Yao, A. C.:
{Monotone bipartite properties are evasive}.
{\em SIAM J. Comput.} 17 (1988), 517--520.}
\end{thebibliography}
\end{document}